\documentclass[prl, twocolumn, superscriptaddress, preprintnumbers,amsmath,amssymb,floatfix]{revtex4-1}
\usepackage{graphicx}
\usepackage{epstopdf}
\usepackage{subfigure}

\newcommand{\fig}[1]{Fig.~\ref{#1}}

\begin{document}
\title{Fault tolerant quantum computation with very high threshold for loss errors}
\author{Sean D. Barrett} \email[]{seandbarrett@gmail.com}
\affiliation{Centre for Quantum Science and Technology,
Macquarie University, NSW 2109
Australia 
}
\affiliation{Blackett Laboratory and Institute for Mathematical Sciences, Imperial College London, Prince
Consort Road, London SW7 2BZ, United Kingdom}
\author{Thomas M. Stace} 
\affiliation{School of Mathematics and Physics, University of
Queensland, Brisbane, QLD 4072, Australia}
\begin{abstract}
Many proposals for fault tolerant quantum computation (FTQC) suffer detectable loss processes. 
Here we show that topological FTQC schemes, which are known to have high error thresholds, are also extremely robust against losses. We  demonstrate that these schemes tolerate loss rates up to 24.9\%, determined by bond percolation on a cubic lattice.  Our numerical results show that these schemes retain good performance when loss and computational errors are simultaneously present.
\end{abstract}
\maketitle

One of the most important achievements of quantum information theory was the discovery of
FTQC: arbitrarily precise, scalable quantum computations can be performed using error-prone components, as long as the error rate is below  a certain threshold  \cite{PhysRevA.52.R2493, Preskill1998}.  
Current thresholds, around  $p_{\mathrm{thres}} \sim 10^{-2}$ per elementary operation  \cite{knill2005,1367-2630-9-6-199}, come tantalisingly close to the experimental state of the art.
Of particular interest is the beautiful proposal developed by Raussendorf, Harrington and coworkers \cite{PhysRevLett.98.190504,raussendorf:062313, Raussendorf20062242,1367-2630-9-6-199}
using ideas from topological quantum computing \cite{kitaev2003ftq} and with only nearest neighbour gates.

Many proposals for quantum computing suffer qubit loss, such as photon loss, atom or ions escaping from traps, or, more generally, the leakage of a qubit out of the computational basis in a multi-level system. However,  such errors can be detected and located without affecting the state of the remaining qubits.  It is therefore expected that appropriately tailored FTQC schemes can tolerate higher rates for loss errors than for unlocateable errors \cite{haselgrove2006trade}. Certain schemes have loss thresholds of $p_{\mathrm{loss}} < 0.5 $ \cite{varnava:120501, Knill2005detected}, although it is unclear how they  perform in the presence of computational errors (unlocated   errors acting within the computational subspace). 
FTQC proposals 
tolerant to both error types have thresholds of $p_{\mathrm{loss}} \lesssim 3 \times 10^{-3}$ and $p_{\mathrm{comp}} \lesssim  10^{-4}$ \cite{dawson:020501,hayes2009fault}.

In this Letter, we describe a FTQC scheme which tolerates both loss and computational errors with very high thresholds. The thresholds are characterized by a contour in $(p_{\mathrm{loss}}, p_{\mathrm{comp}})$ parameter space passing through the points $(0.249,0)$ and $(0,0.0063)$.
This represents an improvement of almost 2 orders of magnitude over earlier results \cite{dawson:020501} \footnote{In contrast to \cite{dawson:020501, hayes2009fault} we assume deterministic gates.}. Our approach requires only that losses are detected at the final readout stage.

Our scheme combines methods from Raussendorf's topological scheme \cite{PhysRevLett.98.190504,raussendorf:062313,1367-2630-9-6-199} and our previous work on loss tolerance in surface codes \cite{PhysRevA.81.022317,stace:200501}. As in the latter, 
 the loss tolerance threshold 
 follows from the bond percolation threshold on the relevant lattice,  
here the  cubic lattice \cite{PhysRevE.57.230}.
To this end, we develop and implement a new classical algorithm to analyse the syndrome in the presence of both loss and logical errors. Monte carlo simulations of the resulting FTQC scheme yield an estimate of the $(p_{\mathrm{loss}}, p_{\mathrm{comp}})$ threshold contour. This contour exhibits a moderate tradeoff between $p_{\mathrm{loss}}$ and $p_{\mathrm{comp}}$, indicating that the code is robust against both kinds of error.

\begin{figure}
\begin{center}
\includegraphics[height=2.2cm]{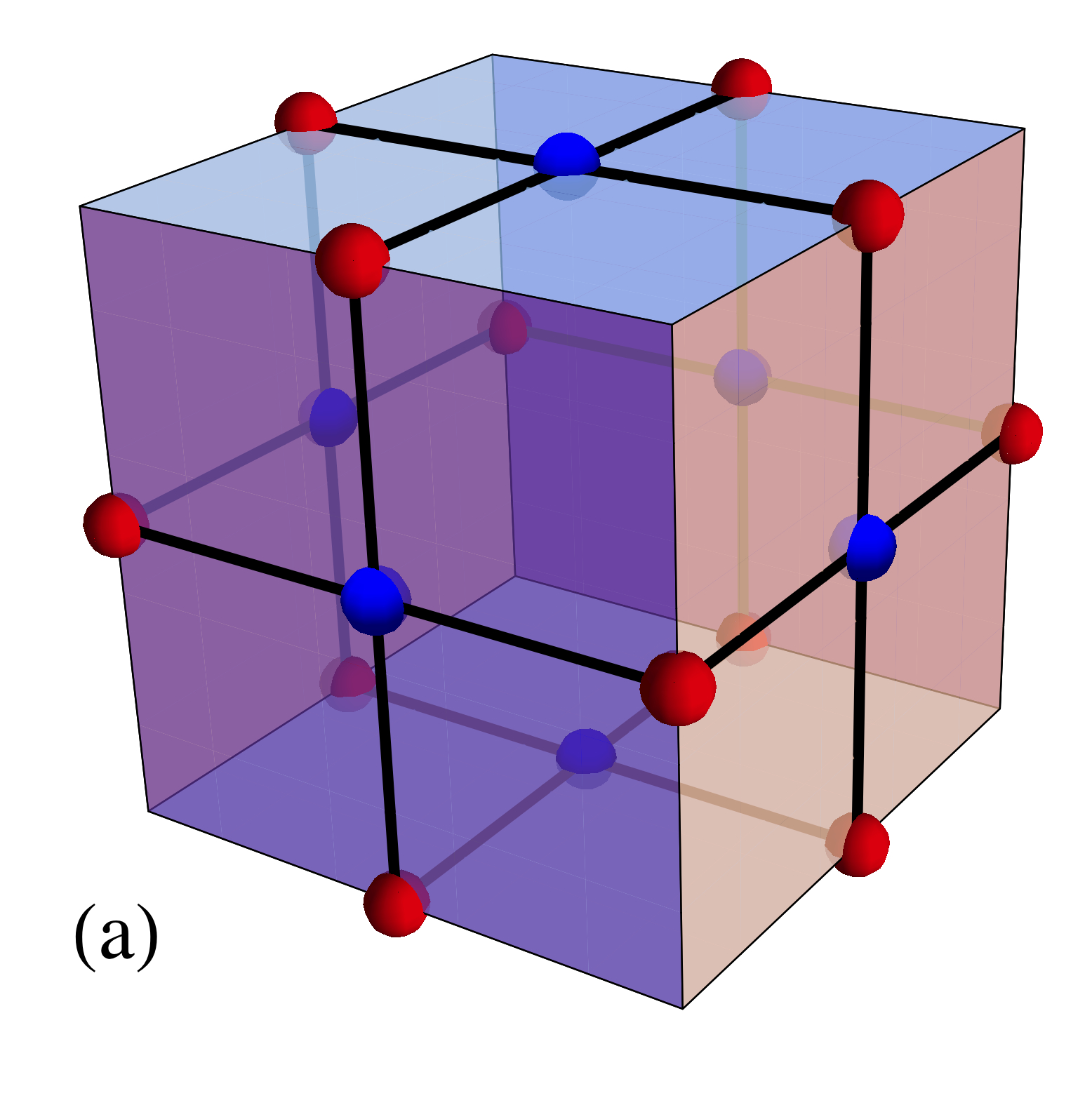}
\includegraphics[width=3cm]{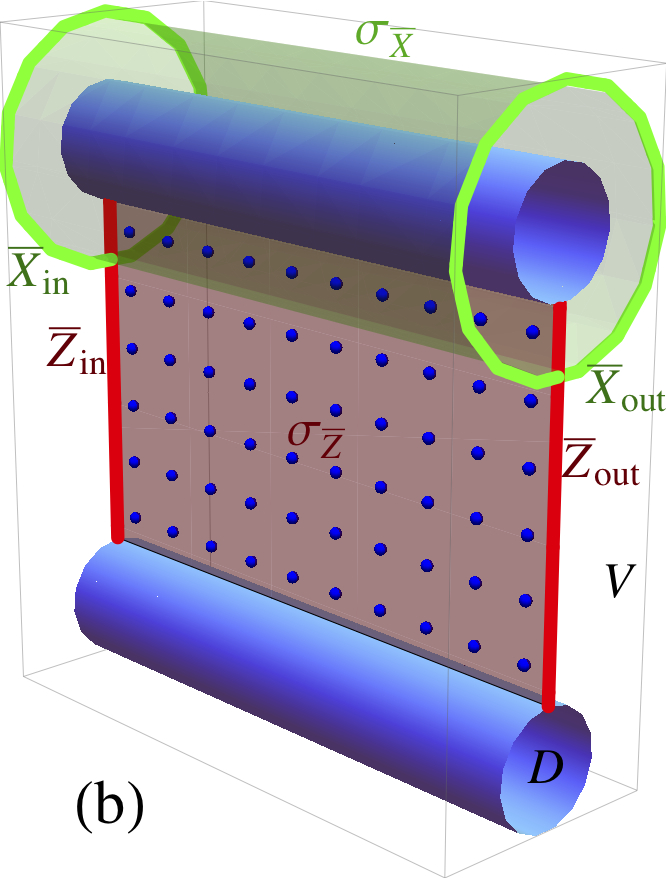}
\includegraphics[width=3cm]{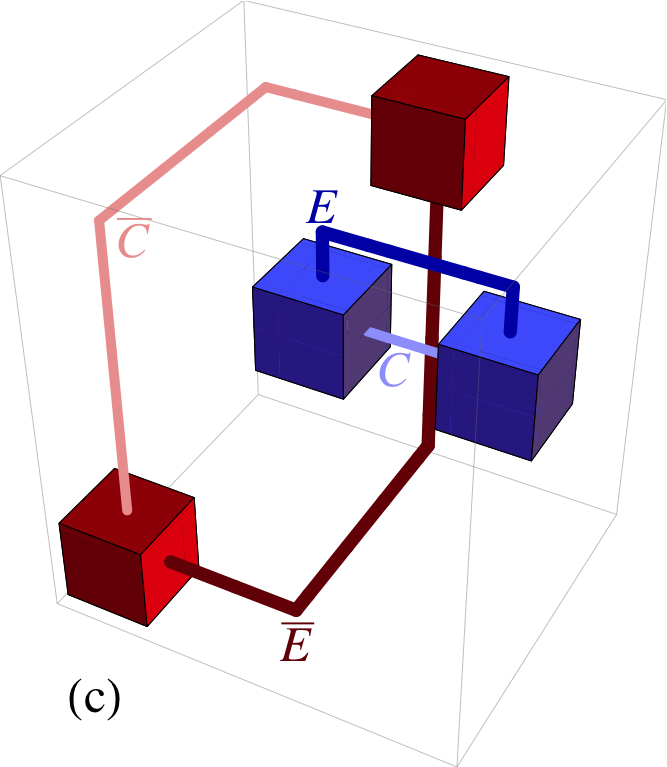}
\caption{\label{fig1} (a) Unit cell showing qubits centered on faces and edges of the primal lattice.  Heavy lines indicate CPHASE gates. (b)
 A primal identity gate showing logical input operators, $\bar X_\mathrm{in}$ and $\bar Z_\mathrm{in}$, and output operators, $\bar X_\mathrm{out}$ and $\bar Z_\mathrm{out}$, and the correlation surfaces, $\sigma_{\bar{X}}$  and $\sigma_{\bar{Z}}$,  that interpolate between them.  Also shown are physical qubits (spheres) living on the correlation surface between the $\bar Z$ operators, and the defect regions, $D$, embedded in the bulk vacuum, $V$. 
  (c) Failed syndromes for error chains on the primal, $E$, and dual, $ E^*$, lattices, and candidate correction chains, $C$ and $C^*$. \label{RaussendorfIllustrationFigure}}
\end{center}
\end{figure}

We first give an overview of Raussendorf's scheme for FTQC. The scheme is based on the one-way quantum computer: that is, a \emph{cluster state} $|C\rangle_\mathcal{L}$, of many physical qubits located on the faces and edges of a cubic primal lattice $\mathcal{L}$, is prepared by first preparing every physical qubit in the state $|+\rangle$, and then applying CPHASE gates between neighbouring qubits on the lattice [see \fig{fig1}(a)]. 
It is convenient to introduce the dual lattice, $\mathcal{L}^*$, whose vertices, edges, faces, and cubes correspond, respectively, to the cubes, faces, edges, and vertices of ${\mathcal{L}}$. Physical qubits reside only on the faces and edges of these lattices. 
Computation proceeds by making a sequence of single qubit measurements on $|C\rangle_\mathcal{L}$, aided by classical processing of the measurement outcomes. Correlations between the measurement outcomes give rise to fault tolerance.

$\mathcal{L}$ is subdivided into three regions, denoted $V$, $D$ and $S$. Qubits in $V$ and $D$ are measured in the $X$ and $Z$ bases, respectively. Qubits in $S$ are measured in either the $Y$ or $(X+Y)/\sqrt{2}$ basis. $D$ comprises a collection of thick structures with some characteristic transverse diameter (`\emph{defects}'), embedded in $V$ [\fig{RaussendorfIllustrationFigure}(b)]. Each pair of defects encodes one logical qubit. As the $D$ qubits are measured in the $Z$ basis, one can omit them altogether, which we assume for the remainder of this paper.

 The topology of the braiding of defect regions effect certain Clifford gates between the logical qubits,   whilst measurement results in $V$ provide topologically-protected fault tolerant error correction of the logical  Clifford gates to arbitrarily high accuracy. 
$S$ comprises a  collection of well-separated single qubits, spread out among the defects. The  $S$ qubits are used to introduce noisy encoded \emph{magic states} into the circuit, which can then be distilled using the very accurate logical Clifford gates in $V$ and $D$,  implementing a universal gate set \cite{bravyi2005universal}.

Magic state distillation tolerates a rather large amount of noise, and so it is found that the thresholds for both loss and computational errors are set by the corresponding thresholds in the bulk topological region, $V$ and $D$. We therefore discuss error correction in $V$ and $D$ in more detail, focusing on the implementation of a fault tolerant identity gate. The reader is referred to \cite{1367-2630-9-6-199,FowlerReview} for details.

The layout of the identity gate is show in \fig{fig1}(b). The boundaries of $D$ are positioned such that faces of $\mathcal{L}$ lie just inside the defect regions. 
The alignment of the defects specify a `simulated time' axis, such that the  identity gate maps the logical qubit from the input (rear) plane, $I$, to the output (front) plane, separated by an integer number of unit cells. Within these planes, the qubit is encoded in a \emph{surface code} \cite{kitaev2003ftq}, containing two holes, which coincide with the intersection of the defect regions with the corresponding plane. A single logical qubit is associated to the pair of holes, with encoded logical operators $\bar{X}$ and $\bar{Z}$ as indicated on \fig{fig1}(b).

To demonstrate the operation of the noiseless gate, it suffices to show  that the input logical operators $\bar{X}_{\mathrm{in}}$, $\bar{Z}_{\mathrm{in}}$ are mapped to the output operators $\bar{X}_{\mathrm{out}}$, $\bar{Z}_{\mathrm{out}}$, under the action of the CPHASE gates and single qubit measurements 
 \cite{gottesman1998heisenberg}.
To this end, we introduce the cluster stabilizer operators, \mbox{$K_f = X_f \bigotimes_{e\in \partial f} Z_e$}, associated with each face $f$ of $\mathcal{L}$ or $\mathcal{L}^*$, where $\partial f$ denotes the qubits at the edges of $f$. 
The initial state of the cluster satisfies $K_f|C\rangle_\mathcal{L} = |C\rangle_\mathcal{L}$. 
Consider the surface $\sigma_{\bar{Z}}$ illustrated in \fig{fig1}(b), consisting of faces in $\mathcal{L}$. The product of the cluster stabilisers on $\sigma_{\bar{Z}}$ yields  $K(\sigma_{\bar{Z}}) = \bigotimes_{f \in  \sigma_{\bar{Z}} } X_f \bigotimes_{e\in \partial \sigma_{\bar{Z}}} Z_e $, where $\partial \sigma_{\bar{Z}}$ denotes the qubits at the boundary of $\sigma_{\bar{Z}}$.  Since \mbox{$\bigotimes_{e\in \partial \sigma_{\bar{Z}}} Z_e=\bar{Z}_{\mathrm{in}} \otimes \bar{Z}_{\mathrm{out}}$},
 it follows that measurements in the bulk effects the operator mapping  $\bar{Z}_{\mathrm{in}} \to \pm\bar{Z}_{\mathrm{in}} \otimes \bar{Z}_{\mathrm{out}}$, where the sign is given by the parity of the measurement outcomes $ \Pi_{f \in  \sigma_{\bar{Z}} } x_f $.

Similarly, the product of cluster stabilisers acting on the surface  $\sigma_{\bar{X}}$, consisting of faces in $\mathcal{L}^*$, yields $K(\sigma_{\bar{X}}) = \bigotimes_{f^* \in  \sigma_{\bar{X}} } X_{f^*} \bigotimes_{{e^*}\in \partial \sigma_{\bar{X}}} X_{e^*} $, and since \mbox{$\bigotimes_{{e^*}\in \partial \sigma_{\bar{X}}} X_{e^*}=\bar{X}_{\mathrm{in}} \otimes \bar{X}_{\mathrm{out}}$} it follows that measurements on $\sigma_{\bar{X}}$ effect the mapping $\bar{X}_{\mathrm{in}} \to \pm\bar{X}_{\mathrm{in}} \otimes \bar{X}_{\mathrm{out}}$. 
Finally, measuring the qubits in $I$ leads to the desired transformation (up to an unimportant Pauli frame update) $\bar{Z}_{\mathrm{in}} \to \pm \bar{Z}_{\mathrm{out}}$ and $\bar{X}_{\mathrm{in}} \to \pm \bar{X}_{\mathrm{out}}$.

Error syndromes are revealed by correlations of measurement outcomes. 
In particular, considering the product of $K_f$'s centred on the faces of a unit cube $c$ of $\mathcal{L}$ leads to the parity check  $P_c = \Pi_{f\in \partial c} x_f$.
 Absent any errors, the parity checks satisfy $P_c=+1$. 
 Failed parity checks, i.e. cubes which instead satisfy 
 $P_c = -1$, reveal the end points of \emph{chains}, $E$, 
  of $Z$ errors (see \fig{fig1}(c)). These chains reside on edges of $\mathcal{L}^*$, and are corrected by inverting the recorded measurement outcomes on some correction chain $C$ (i.e.\ $x_f\rightarrow -x_f$ for $f\in C$) sharing the same endpoints as $E$ (i.e.\ $\partial C=\partial E$). This correction process works provided the combined chain $E+C$ neither winds around a defect, or joins two defects together.

Finding a suitable correction chain  $C$ requires an algorithm for pairing failed parity checks, then finding a suitable path in the lattice between each pair of syndromes.  Choosing $C$ optimally is computationally difficult, since it is equivalent to minimising the free-energy of a particular spin-glass model \cite{dennis:4452,raussendorf:062313}.  However efficient heuristic methods for  decoding error syndromes have been developed, including those using Edmonds'  perfect matching algorithm \cite{wang2003cht,kolmogorov2009blossom}, and belief propagation   \cite{PhysRevLett.104.050504}.

\begin{figure}[!t]
\begin{center}
\includegraphics[width=0.4\columnwidth]{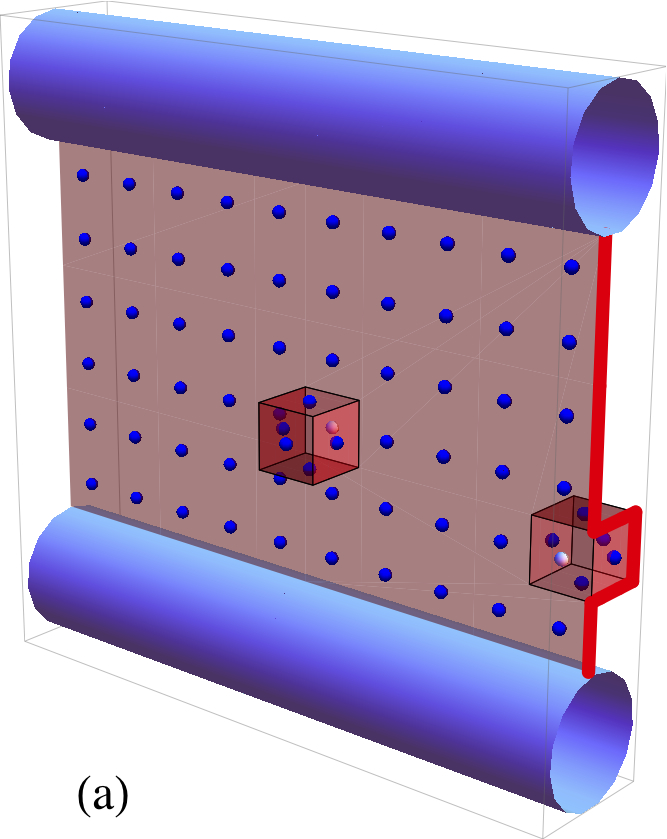}\hfil
\includegraphics[width=0.4\columnwidth]{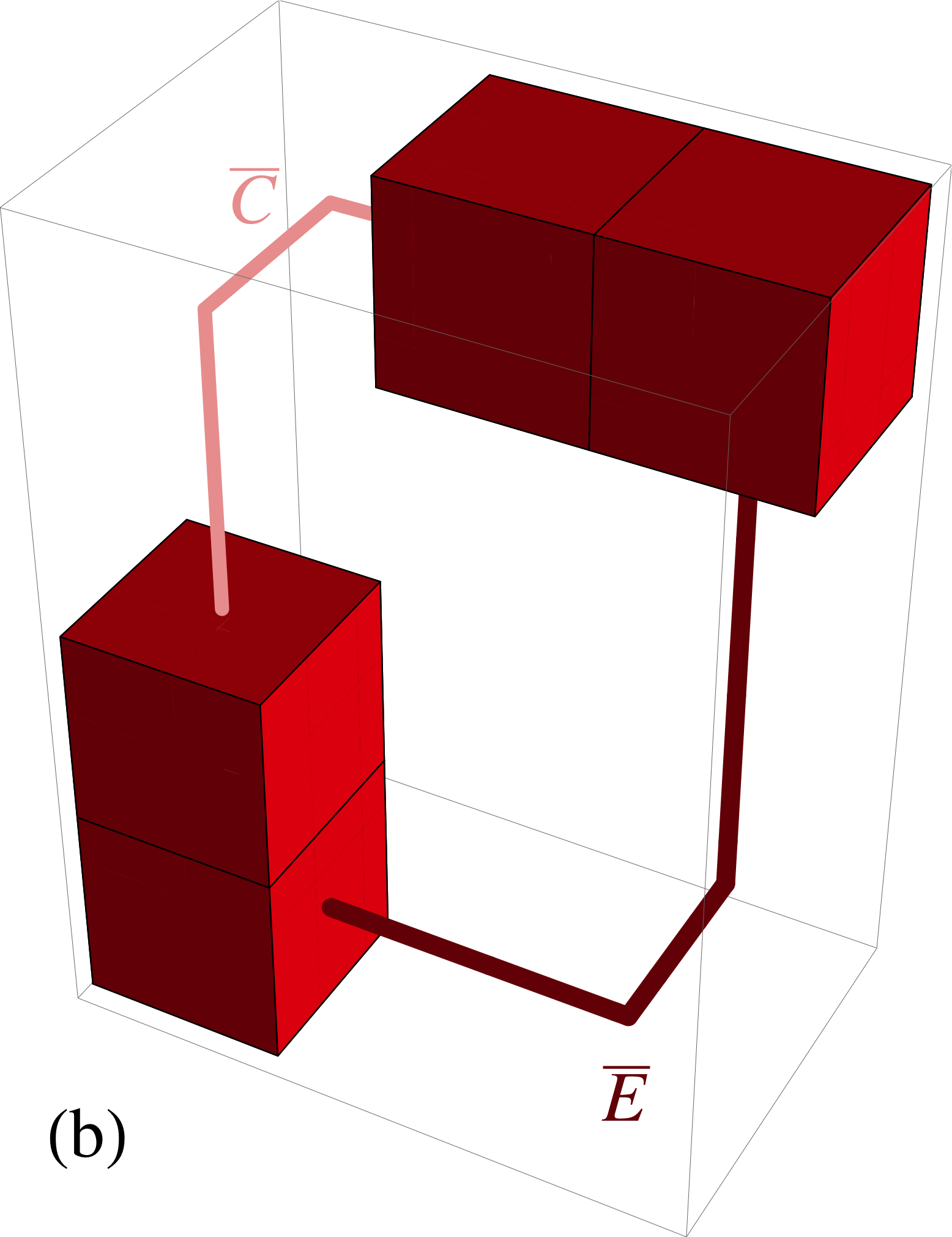} 
\caption{(a) A correlation surface suffering loss is repaired by deforming around lost qubits  (white spheres),  by multiplying by the corresponding check operator.  An affected logical input or output operator is similarly deformed.  (b) In the bulk, super-check operators are formed by products of elementary check operators that are affected by a loss.  Error chains  terminate within a super-check operator. \label{fig2}}
\end{center}
\end{figure}

Realistic error models depend  on the particular implementation under consideration.
As in \cite{Raussendorf20062242}  we assume single qubit depolarising errors take place during preparation, storage, and measurement, with rates $p_P, p_S$ and $p_M$ respectively, whilst  CPHASE gates suffer two-qubit depolarising errors with rate $p_2$.  For numerical calculations, we assume these errors are iid and occur with equal probability,  $p_\mathrm{comp}$, i.e.\ $p_P=p_S=p_M=p_2=p_\mathrm{comp}$. These parameters may vary, with modest changes in the threshold \cite{Raussendorf20062242}.  
For this error model, previous  
results established an error threshold of $p_{\mathrm{comp}} <p_t= 0.0058$ \cite{Raussendorf20062242}.   Note that without losses, exploiting error correlations within each sub-lattice improves the threshold to $p_t = 0.0075$ \cite{PhysRevLett.98.190504}. A further improvement may be possible by exploiting correlations \emph{between} $\mathcal{L}$ and $\mathcal{L^*}$  \cite{PhysRevLett.104.050504}. For simplicity, we do not exploit correlations, at the cost of a marginal reduction in the threshold.

We assume losses are iid with total rate $p_{\mathrm{loss}}$, and only occur either before or after all CPHASE gates have acted upon each qubit, but not at intermediate times. This is consistent with e.g.\  optical lattices where loading losses \cite{PhysRevLett.82.2262} are much higher than storage losses \cite{PhysRevA.77.052309}, or with a photonic implementation where source and detector inneficiencies are expected to dominate. CPHASE gates acting on a lost input are assumed to implement the identity operation, plus depolarizing noise at rate $p_2$, on the remaining qubit. Losses need only be detected at the measurement step, so loss detection is permitted to be a destructive process. Loss detection may be imperfect, but such errors are accounted for in $p_M$ or $p_\mathrm{loss}$.

To deal with these losses, we adapt the Edmonds' matching approach, paralleling \cite{stace:200501,PhysRevA.81.022317}. Suppose qubit $q$ is lost.  The two parity checks $P_{c_q}$ and $P_{c'_q}$ on the adjacent cubes $c_q$ and $c_{q'}$ are incomplete so cannot detect error syndromes.  Likewise any correlation surface $\sigma$ that depends on $q$ is damaged so cannot mediate logical gates. 

We recover from this damage in two ways.  
Firstly,  multiplying $K(\sigma)$ by $K(\partial c_q) = \bigotimes_{f\in \partial c_q} X_f$, yields a new surface $\tilde\sigma$ that avoids $q$ [see \fig{fig2}(a)], i.e. $K(\tilde \sigma)=K(\sigma)K(\partial c_q)$ is independent of $q$.  Then $\tilde \sigma$  determines an equivalent mapping of input operators to output operators. This procedure can be iterated for all of the lost qubits, and succeeds as long as there is some 2D correlation surface that percolates between the logical qubit operators whilst completely avoiding lost qubits.  In the limit where the defect dimensions and spacings between defects becomes large, such a surface can be found providing the loss rate is lower than the cubic-lattice bond percolation threshold, which is approximately $0.249$ \cite{PhysRevE.57.230}.

Secondly, the product of the two damaged check operators $\tilde P_q=P_{c_q}P_{c'_q}$  is independent of $q$, so the sign of $\tilde P_q$  
  yields information about $\partial E$, as illustrated in \fig{fig2}(b).  Thus for each instance of losses, we form new \emph{super}-check operators that are insensitive to the losses.  The lattice of check operators is then no longer cubic, so we perform the syndrome matching on an irregular lattice.

In the context of surface codes, Edmonds' matching has typically been used to find maximum likelihood  correction chains, $C$, that maximise the probability $P(C|\partial E)$ \cite{wang2003cht,1367-2630-9-6-199}.  It is simple to show that such chains are also of minimum length \cite{wang2003cht}.  For a typical pattern of loss, it turns out that there are very many minimum length chains.  Furthermore, as discussed in \cite{PhysRevA.81.022317} some minimum-distance pairings have many more matching chains than other minimum-distance pairings, and so they are \emph{a priori} much more likely. Accounting for this degeneracy in the pairing is computationally cheap, and leads to modest improvements in the threshold. In the 3D FTQC scheme on an irregular lattice, the degeneracy of matchings may also be computed efficiently by modifying Dijkstra's algorithm for finding shortest paths on a graph.

To establish quantitative error thresholds in the presence of losses, we perform Monte Carlo simulations on  $L^3$ cubic lattices with periodic boundary conditions. 
{\parskip 1pt
For each simulation we do the following:
\begin{enumerate}
\itemsep 1pt
\parskip 0pt
\item Generate an instance of errors $E$ and losses.
\item From losses, derive $\tilde P$ and $\tilde \sigma$. If losses percolate the cubic lattice then {\sc{failure}} (no such $\tilde \sigma$ exists).
\item From $E$,  derive $\partial E$ on the irregular lattice.
\item From $\partial E$, compute the correction chain $C$.
\item Compute the homology class $\mathcal{H}_{E+C}$ (count the number of intersections of $E+C$ with $\tilde \sigma$, mod 2).
\item If $\mathcal{H}_{E+C}=0$ then {\sc{success}}.  Otherwise {\sc{failure}}.
\end{enumerate}}

At a given loss rate, we generate $198,000$ instances of errors  whilst varying $p_\mathrm{comp}$ 
and $L=6,8,\ldots 16$.  For each loss rate, we fit a Taylor's expansion of a universal scaling function 
\mbox{$
p_\mathrm{fail}(p_{\mathrm{loss}},p_{\mathrm{comp}},L)\approx a+b\, x+c \,x^2,
$}
where \mbox{$x=(p_\mathrm{comp}-p_t)L^{1/\nu}$}, with fitting parameters $p_t$ (the loss-dependent computational error threshold), $\nu$ (the scaling exponent), $a, b$ and $c$.

\begin{figure}
\begin{center}
\includegraphics[width=\columnwidth]{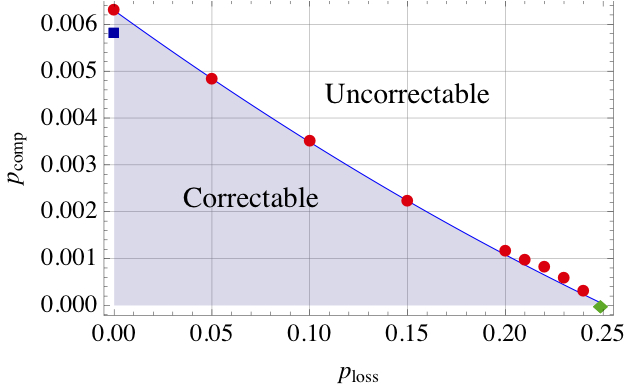}
\caption{Phase diagram showing correctable region of parameter space.  Red circles are numerically calculated computational error thresholds, $p_t$ at different loss rates, for an error model in which $p_P=p_S=p_M=p_2=p_\mathrm{comp}$.  The green diamond is the 3D bond percolation threshold \cite{PhysRevE.57.230}. The blue square is the threshold computed in \cite{Raussendorf20062242}, which ignores degeneracy in the matching algorithm.\label{fig3}}
\end{center}
\end{figure}

\fig{fig3} is the central result in this paper, and shows $p_t$ at different values of $p_\mathrm{loss}$ (red circles).  The shaded region indicates the correctable region of parameter space; if $p_\mathrm{comp}<p_t$, then errors are almost always correctable (in the limit  $L\rightarrow\infty$).  The solid curve is simply a quadratic fit to the computed values of $p_t$ for $0\leq p_\mathrm{loss}\leq 0.15$
\footnote{Non-universal scaling due to finite size effects limit our confidence in the results for $p_\mathrm{loss}>0.2$; see \cite{PhysRevA.81.022317}.}, and extrapolates to  $p_\mathrm{loss}=0.252\pm0.005$ at $p_\mathrm{comp}=0$, which agrees with the percolation threshold $p_\mathrm{perc}=0.249$ \cite{PhysRevE.57.230} (green diamond) on a 3D cubic lattice. 

The loss-tolerant thresholds above rely on properties of the FTQC scheme in the bulk $V$ and $D$ regions, well away from $S$ qubits.  Special attention must be given to the effect of losses affecting $S$ qubits (and also to nearby $V$ and $D$ regions), which  are used to inject encoded \emph{magic states} into the circuit. Computational errors close to the $S$ qubits increases the effective noise on the encoded magic states to $p_S\approx6 p_{\mathrm{comp}}$, but the threshold for magic state purification is so high that the overall threshold is still set by that of the bulk \cite{1367-2630-9-6-199}. This remains true with losses,  and we now describe a scalable \emph{post-selection} method which, 
whilst profligate, demonstrates that losses near $S$ qubits do not limit the thresholds.

If a loss occurs either on or near an $S$ qubit, we discard the corresponding encoded magic state by  measuring  the encoded state, i.e.\ fusing the neighbouring defect strands together. 
This may 
be performed in the bulk region, provided the error rates are below the corresponding bulk thresholds. 
Thus, we require that the only magic state qubits which are injected into the circuit are those 
for which {no losses occur} within a $d\times d\times d$ volume centred on the $S$ qubit itself. Then the post-selected error rate on the encoded magic states is \mbox{$p_S \approx 6 p_{\mathrm{comp}} + p_{\mathrm{fail}}(p_{\mathrm{loss}},p_{\mathrm{comp}},d)$}. 
 Below the bulk error threshold, $p_{\mathrm{fail}}$ is exponentially suppressed in $d$. 
 The magic state distillation threshold $p_S < 1/\sqrt{35}$ \cite{bravyi2005universal} fixes $d$  to be a (small) constant and so the additional overhead for this post-selected scheme, $\sim(1-p_{\mathrm{loss}})^{d^3}$, is independent of the size of the algorithm, ensuring scalability.

In this letter we have established high thresholds FTQC for both computational errors and loss errors.  This result, which sets an important benchmark for experimental implementations, follows from the large redundancy of the surface codes that underpin Raussendorf's FTQC scheme.
This redundancy serves two complementary purposes: to enable deformed correlation surfaces to mediate logical gates between encoded qubits, and to enable expanded parity check operators to identify end points of computational error chains.  

A number of lines of enquiry follow from this work. We assumed deterministic CPHASE gates, so relaxing this  to allow heralded, non-deterministic two-qubit gates in the cluster preparation \cite{PhysRevA.71.060310,benjamin2005optical,barrett2008scalable} is open (but see \cite{li2010fully}).  Also, we expect that losses which occur \emph{during} cluster construction will reduce thresholds further, since these induce additional $Z$ errors on neighbouring qubits.  Since these are located errors, percolation phenomena still determine the loss threshold, so we anticipate this will remain substantially higher than the computational error threshold.
Finally, it is important to develop methods to deal with losses on $S$ qubits with lower overhead than that presented here, perhaps by dynamically routing defect regions and $S$ qubits around any revealed  losses.

SDB was funded by the Royal Society and the CQCT. TMS was funded by the Australian Research Council.  We thank Jim Harrington, Austin Fowler, David Poulin and Andrew Doherty for helpful conversations.

\bibliography{lossy}

\end{document}